\documentclass[pdflatex,sn-mathphys-num]{sn-jnl}

\usepackage{graphicx}%
\usepackage{multirow}%
\usepackage{amsmath,amssymb,amsfonts}%
\usepackage{amsthm}%
\usepackage{mathrsfs}%
\usepackage[title]{appendix}%
\usepackage{xcolor}%
\usepackage{textcomp}%
\usepackage{manyfoot}%
\usepackage{booktabs}%
\usepackage{algorithm}%
\usepackage{algorithmicx}%
\usepackage{algpseudocode}%
\usepackage{listings}%
\usepackage{CJKutf8}
\usepackage{times}
\usepackage{tipa}
\usepackage[T1]{fontenc}

\theoremstyle{thmstyleone}%
\theoremstyle{thmstyletwo}%

\theoremstyle{thmstylethree}%

\begin{document}

\title[Article Title]{DOTA-ME-CS: Daily Oriented Text Audio-Mandarin English-Code Switching Dataset}


\author*[1,2]{\fnm{Yupei} \sur{Li}}\email{yupei.li22@imperial.ac.uk}
\equalcont{These authors contributed equally to this work.}

\author[3]{\fnm{Zifan} \sur{Wei}}\email{zifan.wei@st-andrews.ac.uk}
\equalcont{These authors contributed equally to this work.}

\author[4]{\fnm{Heng} \sur{Yu}}\email{hengyu@mail.nwpu.edu.cn}
\equalcont{These authors contributed equally to this work.}

\author[6]{\fnm{Jiahao} \sur{Xue}}\email{jhxue77@gmail.com}

\author[2]{\fnm{Huichi} \sur{Zhou}}\email{huichi.zhou@imperial.ac.uk}

\author[1,2,5]{\fnm{Björn W.} \sur{Schuller}}\email{bjoern.schuller@imperial.ac.uk}

\affil*[1]{\orgdiv{GLAM Team}, \orgname{Imperial College London}, \orgaddress{\city{London}, \country{UK}}}

\affil[2]{\orgname{Imperial College London}, \orgaddress{\city{London}, \country{UK}}}

\affil[3]{\orgname{University of St Andrews}, \orgaddress{\city{St Andrews}, \country{UK}}}

\affil[4]{\orgname{North China Electric Power University}, \orgaddress{\city{Baoding}, \country{China}}}

\affil[5]{\orgdiv{Chair of Health Informatics (CHI)}, \orgname{Technical University of Munich}, \orgaddress{\city{Munich}, \country{Germany}}}

\affil[6]{\orgdiv{Wuhan University of Bioengineering},
\orgaddress{\city{Wuhan}, \country{China}}}

\abstract{Code-switching, the alternation between two or more languages within communication, poses great challenges for Automatic Speech Recognition (ASR) systems. Existing models and datasets are limited in their ability to effectively handle these challenges. To address this gap and foster progress in code-switching ASR research, we introduce the DOTA-ME-CS: Daily oriented text audio Mandarin-English code-switching dataset, which consists of 18.54 hours of audio data, including 9,300 recordings from 34 participants. To enhance the dataset’s diversity, we apply artificial intelligence (AI) techniques such as AI timbre synthesis, speed variation, and noise addition, thereby increasing the complexity and scalability of the task. The dataset is carefully curated to ensure both diversity and quality, providing a robust resource for researchers addressing the intricacies of bilingual speech recognition with detailed data analysis. We further demonstrate the dataset’s potential in future research. The DOTA-ME-CS dataset, along with accompanying code are shared in the Github\footnote{\url{https://github.com/zifanwei/asr-code-switch}}.}




\maketitle

\section{Introduction}

\begin{figure}[ht]
\vskip 0.2in
\begin{center}
\centerline{\includegraphics[width=\columnwidth]{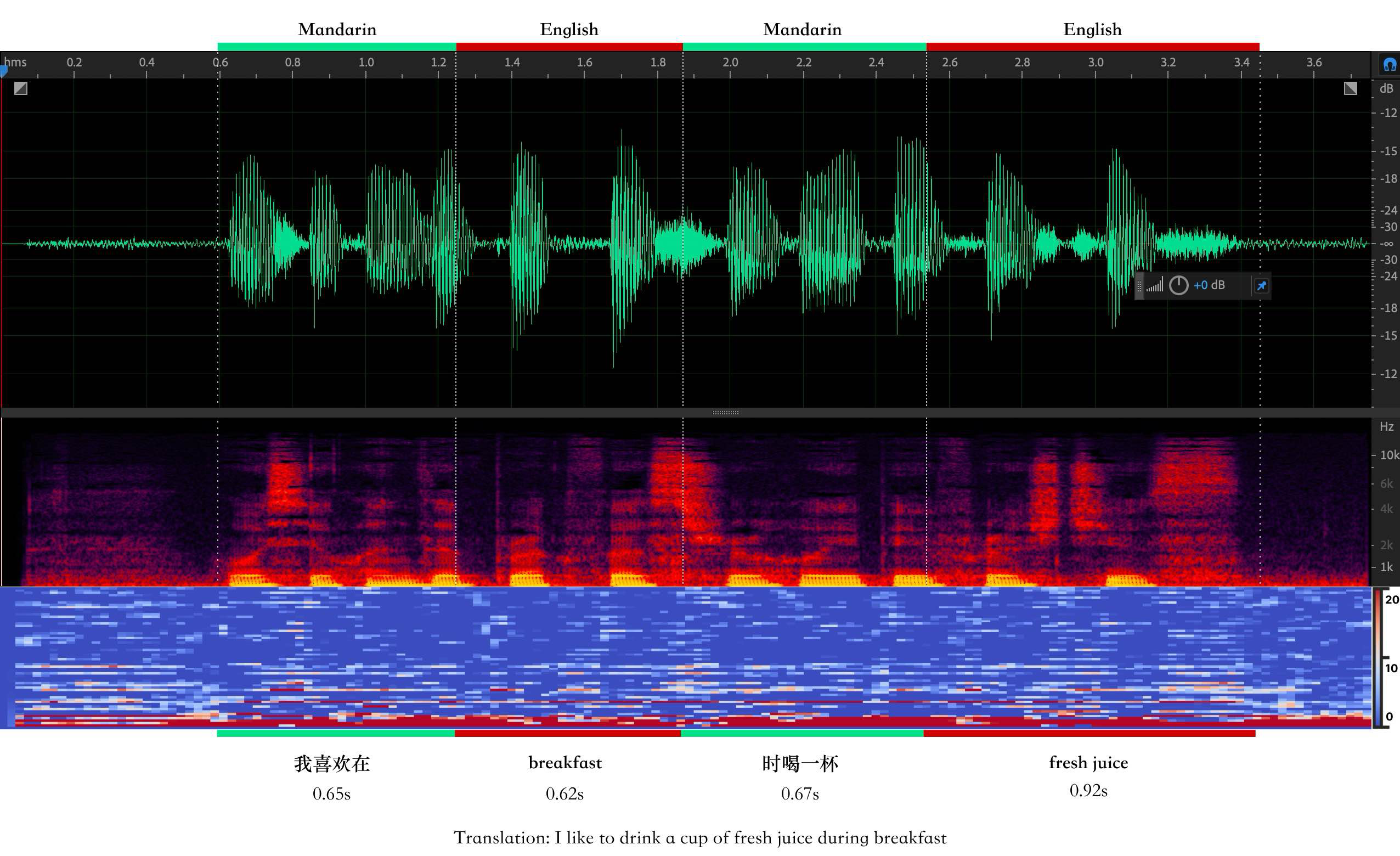}}
\caption{A Mandarin-English Code-switching example, spanning 2.86 seconds. The top green section displays the raw waveform, the middle red section represents the Mel spectrogram, and the bottom section illustrates the Mel-frequency cepstral coefficients (MFCCs) features.}
\label{fig:example}
\end{center}
\vskip -0.2in
\end{figure}

Recent research highlights the significance of code-switching -- the practice of alternating between two or more languages in speech—within multilingual communities \cite{surveycodeswitchinglinguisticsocial, 9074205, article}. The varying contexts and cultural nuances associated with different languages introduce distinct challenges for Automatic Speech Recognition (ASR). One example is illustrated in Figure \ref{fig:example}. Identifying the code-switching point through visual inspection is challenging, although implicit features may be present for models to learn. Another difficulty arises from the differences in language usage habits between code-switching and monolingual speech, making direct translation in the original word order ineffective for understanding the intended meaning.

However, there are multiple advancements in code-switching methodologies recently. Basic techniques such as Connectionist Temporal Classification (CTC)-trained Recurrent Neural Networks have been applied in code-switching tasks \cite{li2019towards}, which, however is inherently limited by the contextual and non-monotonic nature of ASR. Therefore, attention mechanisms \cite{vaswani2017attention} based models such as Wav2Vec 2.0 \cite{baevski2020wav2vec20frameworkselfsupervised}, Sensevoice \cite{an2024funaudiollmvoiceunderstandinggeneration}, and Whisper \cite{radford2022robustspeechrecognitionlargescale} are proposed to better capture dependencies. Detecting switching points remains a crucial, yet challenging approach, as it aligns with the inherent focus of the task. However, current models lack explicit supervision to address this challenge effectively. A common solution are Language Identification (LID) models \cite{liu2023enhancingcodeswitchingspeechrecognition}, which first determine the spoken language before performing ASR, but they are complex due to the highly variable language usage behaviour of speakers. To simplify this process, LLMs have been proposed \cite{xi2024semi}, while they lack interpretability when applied in an end-to-end manner. Despite the progress in this field, existing approaches fail to fully address the aforementioned challenges, indicating that code-switching remains an active and evolving area of research.

High-quality datasets are essential to support the development and evaluation of advanced models. Code-switching has been shown to be a prevalent phenomenon in Mandarin-English speech \cite{wei1995conversational}. However, existing Mandarin-English code-switching datasets, such as SEAME \cite{inproceedings} and ASCEND \cite{lyu10_interspeech}, are derived from spontaneous conversations, making them highly dependent on the performance of the participants. Moreover, these datasets are recorded in controlled, quiet environments, which introduces potential bias, as real-world conversations often occur in diverse scenarios with varying levels of background noise. We have also identified the code\_switch\_yodas\_zh dataset\footnote{https://huggingface.co/datasets/georgechang8/code\_switch\_yodas\_zh}, which consists of voice clips extracted from YouTube videos. However, not all sentences in this dataset involve code-switching, thereby limiting its usability for research focused on code-switching. Datasets from past competitions, such as the ASRU \cite{li2022talcsopensourcemandarinenglishcodeswitching} and TALCS \cite{shi2020asru2019mandarinenglishcodeswitching} datasets, are widely referenced in code-switching research; however, these resources are no longer publicly available. This highlights an urgent need for a comprehensive, publicly accessible dataset to support and drive further research in Mandarin-English code-switching ASR.

To address the existing gaps in available datasets, we propose our own dataset, DOTA-ME-CS: Daily oriented text audio - Mandarin-English - code-switching dataset. This dataset includes annotations for the dominant language within each sentence, alongside ten distinct categories of daily conversational scenes. To ensure the dataset is well-characterised, we provide a detailed descriptive analysis and a comprehensive comparison with existing datasets. To enhance the dataset's realism and diversity, we employ Artificial Intelligence (AI) techniques on a selected subset of our dataset. AI-based timbre synthesis and speed variation are utilised to simulate effects commonly observed in video platforms, and noise addition is to replicate real-world acoustic environments. Furthermore, we conduct essential benchmark evaluations on the dataset to establish baseline performance metrics to not only assess the quality of the dataset but also provide a foundation for further advancements in the field.

Our contributions are as follows:
\begin{itemize}
    \item We introduce a comprehensive Mandarin-English code-switching dataset, enhanced with AI techniques to increase realism. The dataset will be made freely available to the public.
    \item We conduct extensive descriptive analysis of the dataset, providing valuable insights for future research.
    \item We present benchmark evaluations using existing models to establish baseline performance for future studies. 
\end{itemize}

Our paper is organised as follows. In Section \ref{sec2}, we review related works on existing datasets and foundational models in the code-switching domain. Section \ref{sec3} details our data collection methodology, followed by an in-depth analysis presented in Section \ref{sec4}. Benchmark evaluations are provided in Section \ref{sec5}, offering baseline scores for future research. Moreover, we stated impact statement and limitation for our work in Section \ref{sec:add}. Finally, Section \ref{sec6} concludes the paper with a summary of our findings and potential directions for future work.
\section{Related work}
\label{sec2}
\subsection{Datasets}
Several datasets have been developed for the code-switching ASR task, particularly in the Mandarin-English domain. This section provides an overview of these datasets, with a more detailed comparison presented in Section \ref{sec4}.
\paragraph{SEAME}
The SEAME (South East Asia Mandarin-English) \cite{inproceedings} dataset is a conversational speech corpus recorded by a university research team in Southeast Asia. The content primarily consists of transcribed and spontaneous interviews and conversations conducted by students, leading to some randomness in the topics and minor inaccuracies in the transcriptions. The released corpus spans over 190 hours, but portions of the recordings are exclusively in either Mandarin or English. The dataset is publicly accessible through the LDC2015 forum, which requires a membership fee for access.
\paragraph{ASCEND}
The ASCEND (A Speech Corpus for English and Mandarin) dataset \cite{lyu10_interspeech} follows a recording process similar to that of SEAME but involves fewer participants and a shorter total duration of 10.62 hours. This dataset is publicly available through the Hugging Face platform.
\paragraph{ASRU}
The ASRU dataset \cite{shi2020asru2019mandarinenglishcodeswitching}, developed as part of an IEEE challenge, is currently not publicly accessible, limiting its availability for broader research applications. The dataset features a diverse range of participants from China, resulting in a corpus predominantly in Mandarin with fewer English tokens, leading to an imbalance in language representation. Despite this, the code-switching portion of the dataset spans 240 hours.
\paragraph{TALCS}
TALCS \cite{li2022talcsopensourcemandarinenglishcodeswitching} is a dataset derived from a different source but remains inaccessible due to licensing restrictions. It comprises 587 hours of recordings from English-taught sessions within the TAL Education Group, introducing bias as the speakers are predominantly novice English learners, and the content is highly variable and unstructured.

Several additional datasets have been developed. For instance, one extends the aforementioned resources by incorporating data augmentation techniques \cite{wan-etal-2023-new}. Other datasets include the Taiwanese-English code-switching corpus CECOS \cite{6085992}, as well as those based on different language pairs, such as English-Hindi \cite{sreeram2018hindienglishcodeswitchingspeechcorpus}. However, this study primarily focuses on Mandarin-English datasets due to their larger user base, which facilitates more robust analysis. This focus is not a limitation, as models trained on this language pair can be effectively transferred to other languages through fine-tuning, thereby enhancing their applicability in multilingual settings.

\subsection{ASR models}
There exist several important models for ASR, particularly in code-switching, such as CTC-based models \cite{li2019towards}, \cite{naowarat2021reducing}, which focus on mitigating spelling inconsistencies. Additionally, LID detection models \cite{liu2024aligning}, \cite{wang2023language} have been proposed; however, Shen \cite{shen2022does} suggests that the language of individual words has limited impact on recognition accuracy. These models typically either obscure their underlying code or concentrate on addressing very specific, narrow issues.

In light of this, we provide an overview of three baseline deep learning models for ASR in the context of our dataset evaluation. \emph{SenseVoice} \cite{an2024funaudiollmvoiceunderstandinggeneration} is a CTC + attention hybrid encoder with ~200M parameters, trained on the FunAudioLLM multilingual corpus. It integrates ASR with auxiliary tasks including language identification, emotion detection, and audio event detection, with non-autoregressive CTC decoding for fast inference. \emph{Whisper(large-v3)} \cite{radford2022robustspeechrecognitionlargescale} is a Transformer encoder-decoder model with ~1.5B parameters, trained on 680K hours of multilingual data. Its multi-task training (ASR, translation, and language identification) with cross-attention between encoder and decoder ensures high-quality outputs. \emph{Paraformer} \cite{gao2022paraformer} is a non-autoregressive Transformer with 220M parameters, trained on over 16K hours of Mandarin speech. It employs continuous integrate-and-fire (CIF) for alignment-free modeling and is optimized for inference speed. We utilise these three models for code-switching ASR due to their strong capabilities in speech recognition and their adaptability as foundational models in our proposed dataset. Table \ref{tab:asr_models} summarizes the architectural and training details of these three models.

\begin{table}[ht]
\centering
\caption{Architectural and training details of the three pre-trained ASR models used for benchmarking.}
\label{tab:asr_models}
\begin{tabular}{p{2.5cm}p{3.2cm}p{1.1cm}p{2.2cm}p{3.5cm}}
\toprule
\textbf{Model} & \textbf{Architecture} & \textbf{Params} & \textbf{Training Data} & \textbf{Key Notes} \\
\midrule
Whisper-large-v3
& Transformer encoder-decoder
& $\sim$1.5B
& 680K hrs, multilingual
& Multi-task (ASR + translation + LID); cross-attention between encoder and decoder \\
\addlinespace
SenseVoiceSmall
& CTC + attention hybrid encoder
& $\sim$200M
& Multilingual (FunAudioLLM)
& Supports emotion/audio event detection; non-autoregressive CTC decoding \\
\addlinespace
Paraformer
& Non-autoregressive Transformer
& $\sim$220M
& Mandarin, $\geq$16K hrs
& CTC-based; continuous integrate-and-fire (CIF); optimized for inference speed \\
\bottomrule
\end{tabular}
\end{table}

\section{Dataset collection}
\label{sec3}
\begin{figure*}[t!]
\vskip 0.2in
\begin{center}
\centerline{\includegraphics[width=\columnwidth]{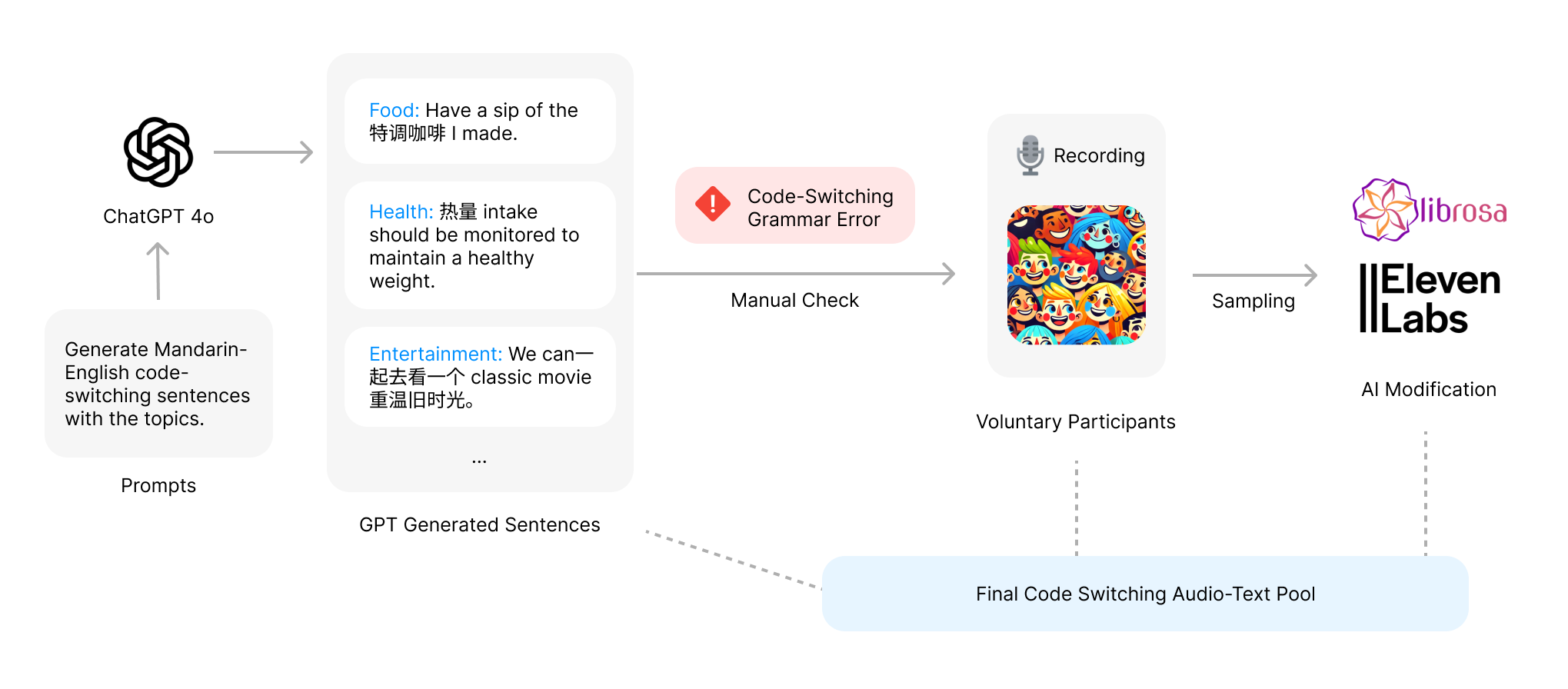}}
\caption{Data collection process of DOTA-ME-CS. Prompts are designed and input into GPT-4o to generate daily-oriented script text. After manually checking for grammatical issues and ensuring the sentences contain code-switching, we invite bilingual volunteers fluent in the switching languages to record their audio. A small subset of the recordings is modified to introduce noise, AI-generated timbres, and speed variations to add diversity and bias to the dataset. Each text-audio pair is uniquely matched to facilitate training for ASR models.}
\label{fig:pipeline}
\end{center}
\vskip -0.2in
\end{figure*}

We introduce a new dataset, the DOTA-ME-CS dataset, designed to address the gaps in existing resources and accelerate the development of code-switching ASR systems. The dataset is carefully curated, with its contents systematically regulated and categorised to align with daily-oriented use cases. Additionally, it has been enhanced to be more comprehensive through AI-driven modifications. The pipeline for dataset creation is illustrated in Figure \ref{fig:pipeline}.

\subsection{Prompts design}

Previous datasets have placed excessive emphasis on scenario realism, often resulting in a lack of flexibility in content creation. To address this limitation, we have developed scripted prompts that participants must follow, ensuring consistency while avoiding freestyle input. This approach eliminates the need for additional annotations, guaranteeing that the audio recordings accurately align with the text labels, which is advantageous for the development of ASR models.

However, we aim to avoid having the scripts deviate too far from everyday communication to meet the specific accuracy demands of ASR models. To achieve this, we employ instruction tuning \cite{zhang2024instructiontuninglargelanguage} to guide LLMs in mimicking natural, day-to-day communication patterns.

Furthermore, we note that many previous datasets predominantly consist of content in a single language, which we believe may not adequately represent the characteristics of code-switching. To maintain balance in language usage within a code-switching scenario, we assign a dominant language in each case. The final prompt is as follows: 
\emph{Generate Mandarin-English code-switching sentences with the $topic$.
Format: id\textbar text\textbar topic, e.g., 201\textbar Shall we try the new \begin{CJK}{UTF8}{gbsn}火锅\end{CJK} restaurant for dinner?\textbar Food.
The given topic cannot be changed.
Duplicated sentences are not allowed.
Sentences are suggested to mimic daily conversational style.
Each sentence must contain more English words than Mandarin, and there must be more than one Mandarin word.}

\subsection{Detailed daily categories}
As demonstrated above, we use the $topic$ to purposefully control the diversity of daily scenarios. To this end, we have selected ten common scenarios: \emph{Education, Entertainment, Environmental Protection, Food, Health, Home, Life, Pets, Travel, Work}, which constitute the complete set of scripts. Furthermore, as shown below, the quality of the generated content improves when more detailed categories are specified in the prompts.

We regard the LLM zero-shot generation process \cite{raffel2023exploringlimitstransferlearning} as a probabilistic model where the output is represented as $P(y|x)$, where $y$ is the generated output and $x$ refers to the conditions provided by the prompts. Assuming that each sub-category is independent and identically distributed (i.i.d.), we compare $P(y|x_i)$ with $P(y|x_1, ..., x_\infty)$, where $x_i$ represents one of the categories. In the case where no specific topics are provided, we consider all topics as potentially relevant, assuming the LLM utilises all the knowledge it possesses. Ultimately, we show that $P(y|x_1, ..., x_\infty)$ is smaller than $P(y|x_i)$, as presented in Equation \ref{eq:iid}.

\begin{align}
    P(y) &= P(y|x_1, \dots, x_\infty) \label{eq:iid} \\
         &= \frac{P(x_1, \dots, x_\infty|y) P(y)}{P(x_1, \dots, x_\infty)} \text{(Bayes' Theorem)} \notag \\
         &= \frac{P(x_i|y) P(x_1, \dots, x_{i-1}, x_{i+1}, \dots, x_\infty|y) P(y)}{P(x_i) P(x_1, \dots, x_{i-1}, x_{i+1}, \dots, x_\infty|y)} \textbf{(i.i.d.)} \notag \\
         &= P(y|x_i) \frac{P(x_1, \dots, x_{i-1}, x_{i+1}, \dots, x_\infty|y)}{P(x_1, \dots, x_{i-1}, x_{i+1}, \dots, x_\infty)} \notag \\
         &< P(y|x_i) \notag
\end{align}

Higher probabilities lead to higher-quality scripts. Additionally, there may be noise and bias in many hidden categories, such as arranged dating with cultural bias, which can intuitively affect performance if the topics are not explicitly specified in the prompt. We could also use the topics to analyse our dataset.

To further demonstrate that our scripts capture the full range of code-switching phenomena, we note two additional dimensions of coverage beyond topical diversity. First, the equal assignment of English-dominant and Mandarin-dominant prompts ensures that both languages serve as the matrix language, capturing both directions of code-switching. Second, the average of 2.22 switching points per utterance (Table \ref{tab:switching}) combined with the diverse POS distribution across both languages (Table \ref{tab:POS tag}) confirms that the scripts contain switching at varied syntactic positions and grammatical categories. The 34 speakers from both China and the UK further ensure that the acoustic realization of these contexts spans multiple English varieties and L2 accents (see Section 3.4 for details).

To assess the naturalness of the GPT-4o-generated code-switching stimuli, we conducted a post-hoc human evaluation with 8 bilingual raters (4 male, 4 female; aged 18–25; 4 English-dominant, 4 Mandarin-dominant). All raters were fluent in both Mandarin and English, with no formal linguistics or phonetics training. Each rater evaluated 100 randomly sampled stimuli on a 5-point Likert scale (1 = highly unnatural; 5 = perfectly natural). The average rating was 4.12 $\pm$ 0.61 (median = 4.0), with moderate inter-rater agreement (Fleiss' coefficient = 0.43). These results indicate that the generated code-switching utterances are perceived as generally natural by bilingual speakers, though we acknowledge that individual stimuli may occasionally exhibit unnaturalness, a known limitation of LLM-based text generation.

\subsection{Long recordings}
To enhance the comprehensiveness of our dataset, we address the limitation of speech samples restricted to approximately 5 seconds, which may not fully represent real-world applications. In practice, many speech instances are longer, posing challenges for current ASR models. Prolonged speech durations can result in a loss of dependencies or critical information, as noted by Koluguri et al.\  \cite{koluguri2023investigatingendtoendasrarchitectures}. To align in this challenge with our dataset, we designed complementary prompts, \emph{Time 10-15 seconds, about 100 words}, instructing LLMs to generate extended scripts. These scripts are tailored to produce speech samples lasting approximately 10 to 15 seconds.
\subsection{Human recording}
We invited 34 college students with academic backgrounds in China and the United Kingdom to participate in voice recordings based on scripts provided to them. All participants were required to be fluent in both English and Mandarin. Approximately 18 participants were based in the UK (primarily at Imperial College London, with 1–5 years of residence) and acquired English through academic immersion; the remaining 16 were based in China and learned English through formal education. Participants self-assessed their language dominance via a pre-recording questionnaire: 12 reported English dominance, 18 reported Mandarin dominance, and 4 identified as balanced bilinguals. None of the participants had professional voice training or formal music education, ensuring naturalistic speech production. Each participant was allowed to record up to 300 scripts, covering both English-dominant and Mandarin-dominant utterances. Participants were instructed to record their voices in a quiet environment, ensuring no other human voices were captured. Participation in this study was voluntary, and all participants provided informed consent for the use of their recordings in further research and potential modifications. This research was reviewed and approved by the ethics committee of the institution where the study was conducted, in accordance with established ethical guidelines and standards for research involving human participants.

\paragraph{Recording protocol}
Recordings were made in a quiet indoor setting (dormitory room or office), with participants instructed to ensure no other human voices were present. We did not use a professional acoustic booth; the quiet-environment requirement was verified by experimenters through sample inspection before acceptance. Participants used their own devices, laptops or smartphones with built-in or headset microphones, and standard voice recording applications, exporting the output as 16-bit WAV files. While device heterogeneity introduces some variability, all recordings were required to meet minimum quality standards: audible speech throughout and no waveform clipping. Recordings that failed these checks were rejected and re-recorded. All accepted recordings were post-processed with peak normalization to 3 dBFS to ensure consistent loudness across participants. We acknowledge that device variability and the absence of a controlled acoustic booth are limitations of this real-world recording protocol; however, the post-hoc physical feature analysis (Table \ref{tab:physical}) confirms reasonable consistency in sampling rate, amplitude, and sample width across the dataset.
\subsection{AI modification}
To ensure high-quality human voice recordings, participants were required to record in a quiet environment and at a normal speaking pace. However, this approach sacrifices some realism, as real-world conversations often occur in more complex scenarios. Many everyday interactions are influenced by background noise and may deviate from calm delivery due to emergencies or emotional factors. To address this limitation, we employed AI tools to simulate such conditions by introducing background noise and varying the speech speed in selected samples.

We randomly selected a subset of recordings and used Librosa \cite{mcfee2015librosa} to modify the playback speed. Speeds were adjusted to 0.75x, 0.5x, 1.25x, 1.5x, and 2x, with 200 audio recordings modified for each speed.
Additionally, we incorporated five types of background noise—\emph{highways, war, natural noise, white noise, and playground sounds}\footnote{https://pixabay.com/sound-effects/}. However, due to the Lombard effect \cite{brumm2009lombard}, which suggests that individuals tend to increase their vocal intensity in the presence of background noise, we adjust the noise pitch to 0.8× to balance it with the original speech and create more realistic scenarios. This adjustment results in 200 recordings per noise type.

Moreover, certain deepfake audio techniques employ voice conversion (VC) \cite{app13053100} to alter the timbre, and other speech processing may need to meet privacy protection requirements to conceal the original timbre \cite{li2024safeear}. To simulate these scenarios, we used AI-generated timbres\footnote{https://elevenlabs.io/app/voice-library} to replace the voices in some samples, utilising Librosa for the conversion. Specifically, we selected five types of AI-generated timbres, including two male and three female voices, and created 200 recordings for each timbre.

\section{Data analysis}
\label{sec4}
\subsection{Audio Analysis}
\paragraph{Duration length}
The duration statistics are detailed in Table \ref{tab:duration}. The notation ``N/A" indicates that the information was either not disclosed in the corresponding paper or the dataset was inaccessible.

\begin{table}[ht]
\centering
\begin{tabular*}{0.95\textwidth}{@{\extracolsep{\fill}}lccccc}
\toprule
\multicolumn{2}{c}{\textbf{Dataset}} & \textbf{Total (Hours)} & \textbf{Avg (Sec)} & \textbf{Var} & \textbf{\#Samples} \\ 
\midrule

\multirow{3}{*}{Ours} 
    & Long     & 6.29   & 15.08 & 14.91 & 1,501  \\
    & Short    & 12.25  & 5.66  & 2.50  & 7,799  \\
    & Overall  & 18.54  & 7.18  & 16.51 & 9,300  \\ 
\midrule
\multicolumn{2}{l}{SEAME}   & 30.00   & 4.30  & N/A  & 25,123   \\
\multicolumn{2}{l}{ASCEND}  & 10.62   & 3.10  & N/A  & 12,314   \\
\multicolumn{2}{l}{ASRU}    & 240.00  & N/A   & N/A  & N/A      \\
\multicolumn{2}{l}{TALCS}   & 587.50  & 5.72  & N/A  & 370,000  \\ 
\bottomrule
\end{tabular*}
\caption{Comparison of dataset duration}
\label{tab:duration}
\end{table}

It is evident that, among the currently accessible datasets, our dataset contains a competitive number of exclusively code-switching samples. Additionally, we include long scripts, which are absent in other datasets. Furthermore, our dataset features a competitive number of samples, as other datasets often contain a great proportion of monolingual samples, whereas ours exclusively consists of code-switching samples.
\paragraph{Phoneme}
Phonemes serve as fundamental acoustic-phonetic features, and their distribution in a dataset affects the performance of phoneme-based speech recognition models. We utilise the International Phonetic Alphabet (IPA) \cite{ipa1999handbook} to represent the phonemes in our dataset. The phonemes identified in the recordings, along with their corresponding frequencies, are presented in Table \ref{tab:phoneme_frequencies}.

\begin{table}[htbp]
\centering
\begin{tabular}{@{}llcc@{}}
\hline
\textbf{Class} & \textbf{Phoneme} & \textbf{Frequency (English)} & \textbf{Frequency (Mandarin)} \\ \hline
\multirow{9}{*}{Plosives}
& p  & 6862  & 3741 \\
& b  & 4147  & -- \\
& t  & 19822 & 11397 \\
& d  & 9854  & -- \\
& k  & 11568 & 6042 \\
& \textipa{g}        & 2468  & -- \\
& \textipa{p\super h} & --    & 1278 \\
& \textipa{t\super h} & --    & 4133 \\
& \textipa{k\super h} & --    & 3338 \\
\hline
\multirow{11}{*}{Fricatives}
& f  & 5441  & 2649 \\
& v  & 5014  & -- \\
& \texttheta          & 988   & -- \\
& \textipa{D}         & 4046  & -- \\
& s  & 14805 & 1176 \\
& z  & 6786  & -- \\
& \textipa{S}         & 2757  & -- \\
& \textipa{Z}         & 123   & -- \\
& h  & 2550  & -- \\
& x  & --    & 5623 \\
& \textipa{\textrtails} & --  & 17397 \\
\hline
\multirow{6}{*}{Affricates}
& \textipa{tS}         & 1603 & -- \\
& \textipa{dZ}         & 1844 & -- \\
& \textipa{ts}         & --   & 4826 \\
& \textipa{ts\super h}  & --   & 1619 \\
& \textipa{tC}         & --   & 8116 \\
& \textipa{tC\super h}  & --   & 3321 \\
\hline
\multirow{3}{*}{Nasals}
& m  & 8112  & 4698 \\
& n  & 17026 & 20966 \\
& \textipa{N}         & 5558  & 19443 \\
\hline
\multirow{4}{*}{Liquids \& Glides}
& l  & 12289 & 5153 \\
& \textipa{\textturnr} & 14054 & -- \\
& j  & 14516 & -- \\
& w  & 8670  & -- \\
\hline
\multirow{17}{*}{Vowels}
& i  & 8893  & 44377 \\
& \textipa{I}         & 17700 & -- \\
& e  & 4846  & 21725 \\
& \textipa{E}         & 9160  & -- \\
& \textipa{\ae}       & 7239  & -- \\
& \textipa{@}         & 21354 & -- \\
& a  & 8143  & 35842 \\
& \textipa{a}         & 4330  & -- \\
& \textipa{O}         & 4703  & -- \\
& o  & 3545  & -- \\
& \textipa{U}         & 1319  & -- \\
& u  & 4583  & 15505 \\
& y  & --    & 4291 \\
& \textipa{2}         & 3872  & -- \\
& \textsyllabic{\textturnr}     & 6220  & -- \\
& uo & --    & 3276 \\
& \textsyllabic{\textturnrrtail} & --    & 5491 \\
\hline
\multirow{2}{*}{Diphthongs}
& iou  & -- & 649 \\
& uei  & -- & 1783 \\
\hline
\end{tabular}
\caption{Frequencies of English and Chinese phonemes represented using the IPA. Phonemes listed with a frequency in only one column are exclusive to the corresponding language.}
\label{tab:phoneme_frequencies}
\end{table}
It is evident that Mandarin tends to exhibit a higher frequency of multisyllabic pronunciations, such as \emph{iou} and \emph{uei}, as well as a greater variety of tones, aspirated consonants (indicated by IPA symbols with \super h), and palatalised sounds. In contrast, English is characterised by a prevalence of single-syllable vowel sounds, such as \emph{\textipa{U}}. Despite these differences, both languages share certain basic phonetic features, such as the sounds \emph{t} and \emph{a}. These observations suggest potential avenues for developing new features in language switching recognition, as the distinct distributions of these phonetic elements are evident both within the languages themselves and in our code-switching dataset.

\paragraph{Physical features}
Additionally, we extract physical features directly from the audio signal. We analyse frame rate, which represents the recording quality, also known as the sampling rate; channels, which indicate the number of separate audio tracks in the recording; amplitude, which measures the intensity of the signal; and sample width, which refers to the number of bytes used to store each audio sample, representing the precision of the recording. These features are summarised in Table \ref{tab:physical}. All results represent the average values computed across all samples.

\begin{table}[t!]
\centering
\resizebox{\linewidth}{!}{%
\begin{tabular}{lccccc}
\toprule
\multicolumn{2}{c}{\textbf{Dataset}} & \textbf{Rate (kHz)} & \textbf{Ch.} & \textbf{Max AMP$^{\mathrm{a}}$} & \textbf{Width (B)} \\
\midrule
\multirow{3}{*}{Ours} 
    & Long     & 48.0  & 1.6 & 20,765 & 2.0 \\
    & Short    & 46.5  & 1.5 & 17,587 & 2.1 \\
    & Overall  & 46.7  & 1.5 & 18,117 & 2.0 \\
\bottomrule
\end{tabular}
}
\vspace{1mm}
\caption{Physical features of DOTA-ME-CS.\newline
\small $^{\mathrm{a}}$ Maximum amplitudes. We filtered 300 anomalies exceeding 32,768 due to noise.}
\label{tab:physical}
\end{table}
The statistics indicate that our recordings were conducted in a quiet environment, ensuring standard high quality. With a frame rate of 46.7\,kHz and a sample width of 2.0\,bytes, the recordings meet typical professional audio standards. The channel value, greater than 1, suggests that the recordings are close to stereo, further supporting their reality as stereo audio provides a more immersive and spatially accurate representation of sound leading to more faithful to real-world auditory perception. Additionally, the maximum amplitude of 18,117 reflects a sufficient dynamic range, free from excessive distortion, implying that the recordings are clear and of high quality for subsequent analysis.

Furthermore, we compute formant frequencies using Linear Predictive Coding (LPC) \cite{markel1976linear} coefficients to capture variations in articulation and pronunciation. The average first, second, and third formant frequencies (F1, F2, and F3) for the entire dataset are 804.02\,Hz, 4419.15\,Hz, and 7549.48\,Hz, respectively. The F1 value indicates the presence of mid-to-low vowels, while the F2 suggests that front vowels dominate the dataset. The F3 value reflects reduced lip rounding and the presence of retroflex consonants. These findings provide evidence of distinct phonetic patterns within our dataset.

\begin{figure}[t!]
\vskip 0.2in
\begin{center}
\centerline{\includegraphics[width=\columnwidth]{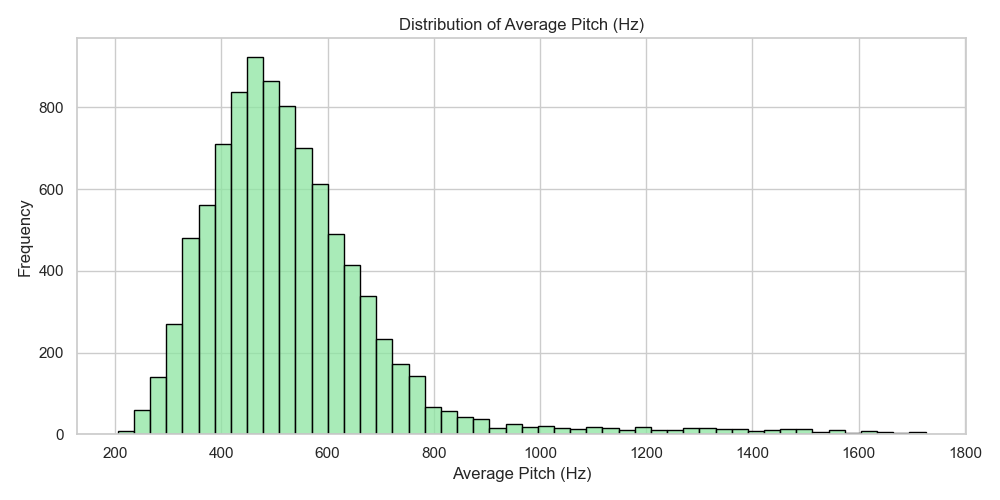}}
\caption{Average pitch of each recording in DOTA-ME-CS}
\label{fig:picth_avg}
\end{center}
\vskip -0.2in
\end{figure}

\begin{figure}[t!]
\vskip 0.2in
\begin{center}
\centerline{\includegraphics[width=\columnwidth]{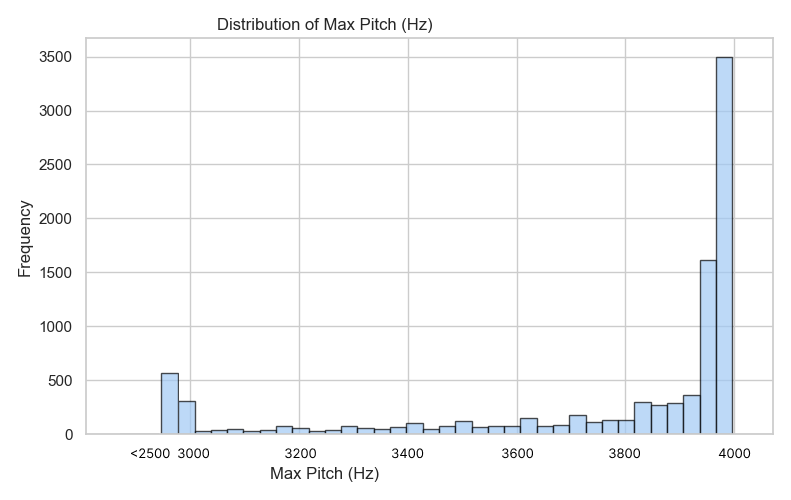}}
\caption{Maximum pitch of each recording in DOTA-ME-CS}
\label{fig:pitch_max}
\end{center}
\vskip -0.2in
\end{figure}

We also present pitch as a key perceptual feature of human auditory processing, serving as a related indicator of the fundamental frequency (F0), which represents the lowest periodic frequency component of a voiced sound. Given the presence of a tonal (Mandarin) and a non-tonal (English) language, F0 should be a key indicator. We note that pitch values are computed across all audio frames without voice activity detection (VAD) preprocessing; consequently, non-speech environmental sounds (e.g., furniture noise, microphone artifacts) present in some recordings may contribute to the pitch statistics, particularly at higher frequencies. The average maximum pitch across recordings is 3617.13 Hz, while the average pitch per recording is 535.04 Hz. The elevated per-recording maxima reflect transient non-speech acoustic events captured in real-world recording conditions, rather than physiological vocal fold vibration. Additionally, we visualise the maximum and average pitch values for each recording in Figures \ref{fig:picth_avg} and \ref{fig:pitch_max}.

The results indicate that a large proportion of pitch values cluster around 500 Hz, with the highest pitch reaching approximately 4000 Hz. This distribution reflects the acoustic diversity inherent in real-world recording conditions, encompassing both speech F0 and environmental contributions. The dataset's recording quality is supported by the physical feature analysis (Table \ref{tab:physical}) and the benchmark results (Table \ref{tab:res}). Moreover, the observed pitch variations serve as important indicators of linguistic boundary marking, phonetic adaptation, and speech rhythm changes, particularly at code-switching points.

We also measured the speaking rate, yielding an average of 2.05 $\pm$ 0.49 words per second. This falls within the typical range of conversational read speech (approximately 2.0–2.5 words per second \cite{yuan2006speakingrate}). The standard deviation of 0.49 indicates natural inter-speaker variability, contributing to the ecological validity of the dataset.

In short, the comparisons above demonstrate the high quality of our dataset and suggest potential directions for improving ASR accuracy by leveraging intrinsic features such as phonemes, amplitudes, pitches, and others, which may indicate many factors such as code-switching points due to the language usage habits of the speakers.

\subsection{Text analysis}
Apart from the audio, we also analyse the scripts. We evaluate the texts to determine whether they represent typical, high-quality, daily-oriented code-switching language. First, we perform Part of Speech (POS) tagging to check the grammar of these scripts, as shown in Table \ref{tab:POS tag}.

\begin{table}[t!]
\centering
\begin{tabular}{ccc}
\hline
\textbf{POS Tag} & \textbf{EN Count} & \textbf{ZH Count} \\ \hline
VERB & 11169 & 19427 \\ 
PRON & 7089  & 5476  \\ 
DET  & 4620  & 810   \\ 
NOUN & 18454 & 20217 \\ 
PART & 1451  & 8356  \\ 
ADP  & 5998  & 2447  \\ 
PROPN & 2719  & 923   \\ 
AUX  & 3039  & -     \\ 
ADV  & 1616  & 6317  \\ 
ADJ  & 6686  & 2197  \\ 
CCONJ & 1437 & 1231  \\ 
SCONJ & 683  & 67    \\ 
INTJ & 43    & 8     \\ 
X    & 9     & 137   \\ 
NUM  & 46    & 2469  \\ 
PUNCT & -    & 1     \\ \hline
\end{tabular}
\caption{POS tag counts for English (EN) and Chinese (ZH)}
\label{tab:POS tag}
\end{table}

As shown in Table \ref{tab:POS tag}, there are notable linguistic and discourse differences between the Mandarin and English corpora. Both languages exhibit a high usage of verbs, nouns, and pronouns; however, English shows a higher frequency of determiners and auxiliary verbs, alongside a lower frequency of particles. These differences could serve as important features when performing ASR in a code-switching task.

We have also evaluated the switching points, presenting the average number of switches per sentence or utterance, along with the average sentence length in tokens, as shown in Table \ref{tab:switching}.

\begin{table}[t!]
\centering
\small 
\begin{tabular*}{\linewidth}{@{\extracolsep{\fill}}lccc}
\toprule
\multicolumn{2}{c}{\textbf{Dataset}} & \textbf{\#Switching Points (avg)} & \textbf{\#Tokens} \\
\midrule
\multirow{3}{*}{Ours} 
    & Short   & 2.02 $\pm$ 0.89  & 11.10 \\
    & Long    & 3.62 $\pm$ 1.31  & 31.73 \\
    & Overall & 2.22 $\pm$ 1.10  & 14.43 \\
\bottomrule
\end{tabular*}
\caption{Average number of switching points and tokens of DOTA-ME-CS.}
\label{tab:switching}
\end{table}

The dataset predominantly contains multiple switching points, with longer recordings exhibiting a higher frequency of switches. This necessitates the development of more accurate ASR systems to effectively handle such complexities.

For statistical purposes, we evaluated the top 10 most frequently used words, excluding stop words, in our dataset, as shown in Table \ref{tab:top10}. These words are commonly used in daily life, indicating that our dataset is representative of everyday language.
\begin{table}[htbp]
\centering
\begin{tabular}{@{}lccc@{}}
\toprule
\textbf{Word (English)} & \textbf{Count} & \textbf{Word (Mandarin)}  & \textbf{Count}\\ \midrule
new & 402 & \begin{CJK}{UTF8}{gbsn}新\end{CJK}&379 \\
let & 306 & \begin{CJK}{UTF8}{gbsn}参加\end{CJK}&333 \\
local & 267 &\begin{CJK}{UTF8}{gbsn}健康\end{CJK}&311 \\
like & 227 &\begin{CJK}{UTF8}{gbsn}学习\end{CJK}& 302\\
love & 210 &\begin{CJK}{UTF8}{gbsn}喜欢\end{CJK}& 298\\
enjoy & 202 &\begin{CJK}{UTF8}{gbsn}生活\end{CJK}&216 \\
good & 166 &\begin{CJK}{UTF8}{gbsn}宠物\end{CJK}& 210\\
food & 157 &\begin{CJK}{UTF8}{gbsn}希望\end{CJK}& 203\\
need & 402 &\begin{CJK}{UTF8}{gbsn}尝试\end{CJK}& 202\\
learning & 151 &\begin{CJK}{UTF8}{gbsn}想\end{CJK}& 195\\
\bottomrule
\end{tabular}
\caption{Top 10 used words in DOTA-ME-CS}
\label{tab:top10}
\end{table}

\section{Experiments}
\label{sec5}
\subsection{Benchmark score}
We have selected Whisper\footnote{https://huggingface.co/openai/whisper-large-v3}, SenseVoice Small\footnote{https://huggingface.co/FunAudioLLM/SenseVoiceSmall}, and Paraformer\footnote{https://huggingface.co/manyeyes/speech\_paraformer-large\_asr\_nat-zh-cn-16k-common-vocab8404-onnx}{} -- widely recognised models in ASR applications for their robust performance -- to assess the quality of our dataset and establish an initial benchmark. To this end, we employed their pre-trained models on our entire dataset without applying any fine-tuning. The evaluation was conducted using Word Error Rate (WER) \cite{morris2004word} and Character Error Rate (CER) as the performance metrics. These metrics reduce the need for extensive data preprocessing, as they inherently ignore case sensitivity and punctuation marks. The results of this evaluation are presented in Table \ref{tab:res}.

\begin{table}[htbp]
\centering
\begin{tabular}{lccc}
\hline
Dataset & Model & WER & CER \\
\hline
\multirow{3}{*}{DOTA-ME-CS} & Sensevoice Small & 0.179 & 0.176 \\
                            & Whisper & 0.206 & 0.213 \\
                            & Paraformer & 0.177 & 0.185 \\
\hline
\multirow{3}{*}{ASCEND} & Sensevoice Small & 0.114 & 0.120 \\
                             & Whisper & 0.289 & 0.290 \\
                             & Paraformer & 0.424 & 0.431 \\
\hline
\end{tabular}
\caption{Benchmark score of three fundamental models on the DOTA-ME-CS and comparison for ASCEND}
\label{tab:res}
\end{table}

The results indicate that Paraformer and SenseVoice achieve superior performance, with Whisper slightly trailing behind. The WER of 0.177 and the CER of 0.176 highlight the high quality of our recordings and scripts. Additionally, these results demonstrate the versatility of our dataset, which supports various model architectures: Paraformer, based on CTC; SenseVoice, a lightweight model utilising an attention mechanism; and Whisper, built upon the transformer architecture. However, the observed differences between CER and WER underscore the code-switching nature of our dataset. Specifically, the results suggest a higher word-level accuracy, while the relatively frequent character-level errors may indicate challenges in accurately capturing specific tokens or phonemes.

Additionally, the comparison between the our dataset and the ASCEND reveals a noticeable performance difference, suggesting that our dataset is of superior quality for model evaluation. On average, the models performance on DOTA-ME-CS, particularly Sensevoice Small and Paraformer, demonstrate higher error rates compared to those on ASCEND. This suggests that DOTA-ME-CS provides more challenging benchmarks, making it a strong candidate for future research and model development, which positions it as a dataset worth proposing for further use in benchmarking.
\subsection{Case study}
We select cases that makes three of the models fail to recognise to analyse. One case in our dataset is id 1167 shown in Table \ref{tab:case1}.

\begin{table}[t!]
\centering
\resizebox{\linewidth}{!}{%
\begin{tabular}{lp{6cm}} 
\toprule
\textbf{Model} & \textbf{Result} \\
\midrule
Groundtruth & Have you tried the \begin{CJK}{UTF8}{gbsn}菠萝包\end{CJK} from the bakery down the street? \\
Paraformer & have you tried the bullleball from the bucky down the street \\
Whisper & Have you tried the pineapple bun from the bakery down the street? \\
SenseVoiceSmall & have you tried the boobao from the bakey down the street \\
\bottomrule
\end{tabular}
}
\caption{Comparison of ASR result for ID 1167 for the same input across different models.}
\label{tab:case1}
\end{table}
The example exhibits a typical feature of code-switching between Chinese and English, particularly with the embedding of the Chinese noun \begin{CJK}{UTF8}{gbsn}菠萝包\end{CJK} (pineapple bun) within an English sentence. Paraformer and Sensevoice directly recognised as \emph{bullleball} and \emph{boobao}, revealing limitations in recognising code-switching scenarios. Moreover ``bakery" is misrecognised as ``bucky," highlighting the model's inadequacy in handling both acoustic features and contextual semantics. On the other hand, the Whisper model accurately translates ``pineapple bun" but fails to faithfully preserve the mixed-language structure, thereby violating the dataset's intent to maintain the original language characteristics. 

We have also selected the case for comparing AI modification, speed change, timbre change, and noise addition below.
\begin{table*}[t!]
\centering
\resizebox{\linewidth}{!}{%
\begin{tabular}{lp{6cm}p{6cm}}
\hline
\textbf{Model} & \textbf{Result} & \textbf{Result for 2x speed modification} \\
\hline
Groundtruth & 
\begin{CJK}{UTF8}{gbsn}我在\end{CJK} Miami Beach \begin{CJK}{UTF8}{gbsn}感受到了典型的\end{CJK} Florida lifestyle. 
& - \\
Paraformer & 
\begin{CJK}{UTF8}{gbsn}我在\end{CJK} miami beach \begin{CJK}{UTF8}{gbsn}感受到了典型的\end{CJK} that florida lifestyle. 
& \begin{CJK}{UTF8}{gbsn}我在蓝牙笔里显示到的也是的\end{CJK}retmy \\
Whisper & 
\begin{CJK}{UTF8}{gbsn}我在\end{CJK} Miami Beach \begin{CJK}{UTF8}{gbsn}感受到了典型的\end{CJK} Florida lifestyle. 
& \begin{CJK}{UTF8}{gbsn}我在\end{CJK} Miami Beach \begin{CJK}{UTF8}{gbsn}感受到了点心的\end{CJK} Forensic Lifestyle \\
SenseVoiceSmall & 
\begin{CJK}{UTF8}{gbsn}我在\end{CJK} miami beach \begin{CJK}{UTF8}{gbsn}感受到了典型的\end{CJK} florida lifestyle. 
& \begin{CJK}{UTF8}{gbsn}我在\end{CJK} Miamiami \begin{CJK}{UTF8}{gbsn}北线收到了电信的\end{CJK} foreign \begin{CJK}{UTF8}{gbsn}人在\end{CJK} \\
\hline
\end{tabular}
}
\caption{Comparison of the ASR result for ID 177 for the same input across different models.}
\label{tab:case_speed}
\end{table*}

In the case of double-speed audio in Table \ref{tab:case_speed}, Paraformer outputs completely deviate from the intended meaning. Not only is key information lost, but irrelevant errors are also introduced, demonstrating the model's weak robustness in handling accelerated audio. For Whisper, while the recognition of ``Miami Beach" is accurate, ``Florida lifestyle" is misrecognised as ``Forensic Lifestyle," indicating a decline in its ability to process English semantics in fast audio scenarios. SenseVoiceSmall includes great repetition and confusion, such as ``Miamiami." This illustrates the model's limited ability to handle mixed-language scenarios involving both Chinese and English in accelerated audio contexts.

\begin{table*}[t!]
\centering
\resizebox{\linewidth}{!}{%
\begin{tabular}{lp{6cm}p{6cm}}
\hline
\textbf{Model} & \textbf{Result} &\textbf{Result for timbre modification}  \\
\hline
Groundtruth & \begin{CJK}{UTF8}{gbsn}你觉得这个新的\end{CJK} documentary \begin{CJK}{UTF8}{gbsn}的拍摄技巧如何?\end{CJK}& - \\
\hline
Paraformer & \begin{CJK}{UTF8}{gbsn}你觉得这个新的\end{CJK} documentary \begin{CJK}{UTF8}{gbsn}的拍摄技巧如何?\end{CJK} & \begin{CJK}{UTF8}{gbsn}你觉得就是sing the documentary的ptial detail如何\end{CJK} \\
\hline
Whisper & \begin{CJK}{UTF8}{gbsn}你觉得这个新的\end{CJK} documentary \begin{CJK}{UTF8}{gbsn}的拍摄技巧如何\end{CJK}? &\begin{CJK}{UTF8}{gbsn}你觉得JOSING的Document rate拍摄细节如何?\end{CJK}\\
\hline
SenseVoiceSmall & \begin{CJK}{UTF8}{gbsn}你觉得这个新的\end{CJK} documentary \begin{CJK}{UTF8}{gbsn}的拍摄技巧如何?\end{CJK}&  nijojo sing the document read paroru\\
\hline
\end{tabular}}
\caption{Comparison of the ASR result for ID 194 for the same input across different models.}
\label{tab:case timbre}
\end{table*}
In the example in Table \ref{tab:case timbre}, the recognition performance deteriorates under altered timbre compared to normal timbre, particularly in terms of English word spelling and semantic integrity. For instance, after timbre modification, Paraformer exhibits a completely disrupted semantic structure. SenseVoiceSmall generates a large number of meaningless words. Whisper, while partially preserving the meaning, output introduces the incorrect word ``JOSING". In contrast, under normal timbre, all three models perform considerably better in restoring the Ground Truth. Paraformer and SenseVoiceSmall produce nearly perfect results, while Whisper, despite minor discrepancies due to traditional Chinese character settings, maintains overall semantic consistency. 

This comparison highlights that timbre modification weakens the models' ability to process mixed Chinese-English sentences, with a pronounced decline in English recognition accuracy. However, the strong performance under normal timbre demonstrates that our dataset effectively covers and adapts to mixed-language scenarios, enabling models to accurately reconstruct meaning. This underscores the dataset's advantages in diversity and semantic fidelity while also revealing areas for improvement in handling extreme conditions such as timbre variations. Enhancing training data diversity in timbre and optimising model robustness could substantially improve performance under non-standard audio conditions.

\begin{table*}[t!]
\centering
\resizebox{\linewidth}{!}{%
\begin{tabular}{lp{6cm}p{6cm}}
\hline
\textbf{Model} & \textbf{Result}& \textbf{Result for noise addition} \\
\hline
Groundtruth & I love to document my travels through \begin{CJK}{UTF8}{gbsn}社交媒体和博客\end{CJK}. & - \\
\hline
Paraformer & I love to document my travels through \begin{CJK}{UTF8}{gbsn}社交媒体和博客\end{CJK}. & I love to document my travels through \begin{CJK}{UTF8}{gbsn}社交媒体和博客\end{CJK}. \\
\hline
Whisper & I love to document my travels through social media and blog. &\begin{CJK}{UTF8}{gbsn}我喜欢通过社交媒体和博客去读书\end{CJK} \\
\hline
SenseVoiceSmall &I love to document my travels through \begin{CJK}{UTF8}{gbsn}社交媒体和博客\end{CJK}. & I love to document my travels through \begin{CJK}{UTF8}{gbsn}社交媒体和博客\end{CJK}.\\
\hline
\end{tabular}}
\caption{Comparison of the ASR result for ID 446 for the same input across different models.}
\label{tab:case noise}
\end{table*}

For case in Table \ref{tab:case noise}, in contrast under noise-free conditions, Paraformer and SenseVoiceSmall produce outputs that are consistent with the Ground Truth, whereas Whisper exhibits a tendency toward automatic translation. This indicates that the Whisper model has a clear translation bias in mixed Chinese-English contexts. Furthermore, under noisy conditions, it introduces additional semantic errors, revealing its limited adaptability to mixed-language sentences and its weaker robustness to noise.
\


In these three cases, all the models perform well on the original examples but exhibit degradation in performance for the modified cases. They deviate substantially from the intended semantics, not only losing key information but also introducing irrelevant error content. This highlights the models' lack of robustness when handling accelerated audio. Additionally, some words are misspelled, further indicating the models' weaknesses. However, such issues are common in real-life scenarios, emphasizing the need for more robust models capable of handling these challenges effectively.

Our dataset demonstrates great advantages in design. It provides high diversity and realistic representation of mixed-language scenarios, helping models learn pronunciation features unique to code-switching. Additionally, the dataset emphasises semantic consistency and the fidelity of Mandarin-English language switching, ensuring that models can not only recognise meaning accurately but also preserve the sentence's original language structure. This example highlights the clear deficiencies of existing models in processing code-switching and specific terms, while our dataset, by covering a wide range of real-world scenarios, enhances a model's ability to adapt to mixed-language contexts, offering substantial support for improving the overall performance of ASR systems.

\subsection{AI modification analysis}
Additionally, we perform the experiments in every subset of our modification strategy, shown in Table \ref{tab:wer_cer_multirow}
\begin{table*}
\centering
\small
\begin{tabular}{ccccc}
\hline
\multirow{2}{*}{\textbf{Model}} & \multirow{2}{*}{\textbf{Condition Type}} & \textbf{Condition} & \multirow{2}{*}{\textbf{Average WER}} & \multirow{2}{*}{\textbf{Average CER}} \\
 &  & \textbf{(e.g., 0.5x, Fin)} &  &  \\
\hline
\multirow{15}{*}{Paraformer} & \multirow{5}{*}{Change speed} & 0.5x & 0.2540 & 0.2593 \\
                             &                              & 0.75x & 0.2246 & 0.2352 \\
                             &                              & 1.25x & 0.2143 & 0.2308 \\
                             &                              & 1.5x & 0.2791 & 0.2924 \\
                             &                              & 2.0x & 0.4400 & 0.4531 \\
                             & \multirow{5}{*}{Change voice} & Fin & 0.2607 & 0.2589 \\
                             &                              & George & 0.2249 & 0.2295 \\
                             &                              & Gigi & 0.2002 & 0.2064 \\
                             &                              & Lily & 0.2393 & 0.2423 \\
                             &                              & Sarah & 0.2139 & 0.2215 \\ 
                             & \multirow{5}{*}{Background}   & Modern-war & 0.1823 & 0.1897 \\
                             &                              & Playground & 0.1908 & 0.2028 \\
                             &                              & Rainy-day & 0.1808 & 0.1956 \\
                             &                              & White-noise & 0.2105 & 0.2160 \\
                             &                              & Road & 0.2060 & 0.2120 \\
                             &                              & General & 0.1770 & 0.1850 \\
\hline
\multirow{15}{*}{SenseVoiceSmall} & \multirow{5}{*}{Change speed} & 0.5x & 0.2467 & 0.2364 \\
                                  &                              & 0.75x & 0.2051 & 0.1972 \\
                                  &                              & 1.25x & 0.1896 & 0.1838 \\
                                  &                              & 1.5x & 0.2218 & 0.2146 \\
                                  &                              & 2.0x & 0.3387 & 0.3285 \\
                                  & \multirow{5}{*}{Change voice} & Fin & 0.3054 & 0.2946 \\
                                  &                              & George & 0.3135 & 0.2999 \\
                                  &                              & Gigi & 0.2750 & 0.2660 \\
                                  &                              & Lily & 0.2621 & 0.2533 \\
                                  &                              & Sarah & 0.3025 & 0.2929 \\
                                  & \multirow{5}{*}{Background}   & Modern-war & 0.1875 & 0.1764 \\
                                  &                              & Playground & 0.2008 & 0.1939 \\
                                  &                              & Rainy-day & 0.2039 & 0.1981 \\
                                  &                              & Road & 0.2149 & 0.2029 \\
                                  &                              & White-noise & 0.2082 & 0.2013 \\
                                  &                              & General & 0.1790 & 0.1757 \\
\hline
\multirow{15}{*}{Whisper} & \multirow{5}{*}{Change speed} & 0.5x & 0.2564 & 0.2606 \\
                          &                              & 0.75x & 0.2394 & 0.2461 \\
                          &                              & 1.25x & 0.2583 & 0.2637 \\
                          &                              & 1.5x & 0.2973 & 0.2970 \\
                          &                              & 2.0x & 0.4186 & 0.4158 \\
                          & \multirow{5}{*}{Change voice} & Fin & 0.2624 & 0.2638 \\
                          &                              & George & 0.2545 & 0.2542 \\
                          &                              & Gigi & 0.2025 & 0.2080 \\
                          &                              & Lily & 0.2130 & 0.2176 \\
                          &                              & Sarah & 0.2368 & 0.2404 \\
                          & \multirow{5}{*}{Background}   & Modern-war & 0.2476 & 0.2546 \\
                          &                              & Playground & 0.2960 & 0.3080 \\
                          &                              & Rainy-day & 0.2889 & 0.2997 \\
                          &                              & Road & 0.2687 & 0.2763 \\
                          &                              & White-noise & 0.2385 & 0.2456 \\
                          &                              & General & 0.2055 & 0.2132 \\

\hline
\end{tabular}
\caption{Average Word Error Rate (WER) and Character Error Rate (CER) for different conditions across three ASR models.}
\label{tab:wer_cer_multirow}
\end{table*}
The results show that the effect of changing speed on the performance of the models shows a distinct trend across all three models for Paraformer, SenseVoiceSmall, Whisper. For Paraformer, WER increases as the speed increases, indicating degraded performance at higher speeds, with the worst results observed at 2.0×. Similarly, SenseVoiceSmall shows an overall unstable trend, but also tends to perform worse under faster speeds, especially at 2.0× where the highest WER is observed. Whisper exhibits a comparable pattern, where WER generally increases with speed and peaks at 2.0×, indicating the poorest performance. Overall, all three models show their worst performance at 2.0× speed, suggesting that excessive speed-up negatively impacts recognition accuracy. For changing voice timbre, Paraformer shows an overall decline in performance as the voice changes, with ``Fin" being the highest performer, and ``Gigi" showing the lowest scores. This suggests that certain voices, may be more challenging for the model to process effectively, possibly due to the model's limitations in handling specific voice characteristics. In contrast, SenseVoiceSmall exhibits more consistent performances. These findings indicate that SenseVoiceSmall and Whisper are more robust to voice changes than Paraformer, which might struggle with specific voice characteristics.

The influence of background noise or conditions on the models shows a more pronounced variation. For Paraformer, the background noise conditions generally lead to a decrease in performance, with "General" background producing the lowest scores while the "Modern-war" and "Road" backgrounds show relatively higher performance. This suggests that some background noises, particularly those that are more complex or varied, may have a more detrimental effect on performance than others. SenseVoiceSmall also shows a decrease in performance with background noise. Interestingly, Whisper demonstrates a stronger ability to handle background noise, which could indicate that it is more resilient in noisy environments compared to the other models. This analysis suggests that while background noise affects all models to some extent, Whisper appears to be the most adaptable to such changes.
\section{Impact Statement and Limitation}
\label{sec:add}
Our paper introduces a human-recorded code-switching dataset, which has profound implications not only for technological development but also for societal uptake. Such a dataset captures the natural linguistic transitions that could be used to develop more reliable ASR models that are better equipped to handle multilingual and code-switched speech. Human-recorded speech bears more salient aspects of natural prosody, pronunciation variation, and speaker-specific characteristics, which contribute to greater generalizability of the model.

From a sociolinguistic point of view, the corpus is a contribution to spoken language diversity which will better ASR performance for those communities where code-switching is a way of life—for multilingual speakers, in general, and in practice. It leverages more availability in automated transcription, voice assistants, and translation systems with less bias against non-monolinguals.

As a matter of ethics, this dataset comes from fully informed speakers who have consented to their recordings being used, and it is guaranteed that speaker privacy and data security are preserved and that such collection met the approval process’s ethics standards. Diverse demographic representation is important to ensure there are no biases in ASR models that could unfairly impact certain minority groups. Proper annotation with an ethical data-sharing policy maintains the goal of fairness and transparency in AI research.

Through furthering the development of technologies in equity, this dataset advances the capacities in ASR and promotes the inclusivity of digital technologies in bridging communication gaps within increasingly globalised societies whereby code-switching has become more and more prevalent within professional, educational, and social interactions.

However, due to limitations in funding, the scope of the dataset scale is constrained, restricting the ability to include a larger and more diverse set of data. However, this limitation does not considerably affect the overall utility of the data, as the datasets used are carefully selected to ensure they remain representative and relevant to the research objectives. Despite the smaller scale, the data has been fully leveraged for its intended purpose, allowing for effective model evaluation and meaningful insights. The focus on quality over quantity ensures that the data still supports the model’s development and testing, maintaining the robustness of the results within the constraints of available resources. As a result, while a larger dataset might offer broader generalization, the current dataset is still highly effective for addressing the research goals at hand.

Moreover, although our scripts are collected by individuals reading pre-written scripts instead of asking people to speak spontaneously, our methodology offers several key advantages. First, we have carefully curated the topics to ensure they closely resemble everyday scenarios, aiming to generate scripts that are highly relatable and reflective of daily life. This thoughtful curation not only enhances the relevance of the data but also improves its utility for the intended research. Additionally, by avoiding sensitive circumstances often encountered during in-person interviews, our approach mitigates potential ethical concerns and ensures a more controlled data collection environment. Furthermore, by relying on scripted readings rather than spontaneous speech, we reduce the likelihood of transcription errors, as the process is more structured and predictable. Overall, our methodology offers a more reliable and efficient approach to data collection, making it particularly valuable in research settings where consistency and accuracy are paramount.

We further note that DOTA-ME-CS contains 9,300 recordings, all of which are exclusively code-switching utterances. In contrast, larger datasets with more total hours (e.g., SEAME at 190 hours, TALCS at 587 hours) include substantial proportions of monolingual content, meaning their effective code-switching sample counts are not proportionally larger. Our emphasis on quality control — including manual script verification, controlled recording protocols, and per-recording quality checks — is an intentional trade-off prioritizing reliability over scale for benchmarking purposes.

\section{Conclusion}
\label{sec6}
We introduced the DOTA-ME-CS dataset, a high-quality, human-recorded code-switching speech corpus enhanced with AI-driven modifications to increase realism. To validate the dataset’s robustness, we conducted extensive data analysis, demonstrating its linguistic richness and overall quality. Additionally, we evaluated benchmark ASR models on DOTA-ME-CS, highlighting existing model limitations and performance gaps. Our findings suggest that leveraging intrinsic features such as phonemes, pitch variations, and part-of-speech (POS) tags can enhance model robustness for ASR in code-switching scenarios. Moving forward, we plan to expand the dataset with real-life conversational data and spontaneous speech recordings. Furthermore, we aim to develop and release more ASR models specifically optimised for code-switching tasks.

\section*{Declarations}

Some journals require declarations to be submitted in a standardised format. Please check the Instructions for Authors of the journal to which you are submitting to see if you need to complete this section. If yes, your manuscript must contain the following sections under the heading `Declarations':

\begin{itemize}
\item Funding: This research was partially supported and funded by the Munich Center for Machine Learning and the Munich Data Science Institute.
\item Conflict of interest/Competing interests (check journal-specific guidelines for which heading to use): Not applicable.
\item Ethics approval and consent to participate: Approved by Imperial College London Ethical Board.
\item Consent for publication: We give consent.
\item Data availability: Codes and data are in the Github\footnote{\url{https://github.com/zifanwei/asr-code-switch}}.
\item Author contribution: Yupei Li: project lead, experiment design and coding, paper writing, data collection. Zifan Wei, Heng Yu: Experiment design and coding, paper writing, data collection. Jiahao Xue: Paper writing, coding, data collection. Huichi Zhou: paper revision. Bj\"orn Schuller: Paper revision, supervison.
\end{itemize}







\bibliography{reference}

\end{document}